\PassOptionsToPackage{unicode}{hyperref}
\PassOptionsToPackage{hyphens}{url}
\PassOptionsToPackage{dvipsnames,svgnames,x11names}{xcolor}
\documentclass[
]{article}
\usepackage{amsmath,amssymb}
\usepackage{lmodern}
\usepackage{iftex}
\ifPDFTeX
  \usepackage[T1]{fontenc}
  \usepackage[utf8]{inputenc}
  \usepackage{textcomp} 
\else 
  \usepackage{unicode-math}
  \defaultfontfeatures{Scale=MatchLowercase}
  \defaultfontfeatures[\rmfamily]{Ligatures=TeX,Scale=1}
\fi
\IfFileExists{upquote.sty}{\usepackage{upquote}}{}
\IfFileExists{microtype.sty}{
  \usepackage[]{microtype}
  \UseMicrotypeSet[protrusion]{basicmath} 
}{}
\makeatletter
\@ifundefined{KOMAClassName}{
  \IfFileExists{parskip.sty}{%
    \usepackage{parskip}
  }{
    \setlength{\parindent}{0pt}
    \setlength{\parskip}{6pt plus 2pt minus 1pt}}
}{
  \KOMAoptions{parskip=half}}
\makeatother
\usepackage{xcolor}
\providecommand{\tightlist}{%
  \setlength{\itemsep}{0pt}\setlength{\parskip}{0pt}}
\NewDocumentCommand\citeproctext{}{}
\NewDocumentCommand\citeproc{mm}{%
  \begingroup\def\citeproctext{#2}\cite{#1}\endgroup}
\makeatletter
 \let\@cite@ofmt\@firstofone
 \def\@biblabel#1{}
 \def\@cite#1#2{{#1\if@tempswa , #2\fi}}
\makeatother
\newlength{\cslhangindent}
\newlength{\csllabelwidth}
\newenvironment{CSLReferences}[2] 
 {\begin{list}{}{%
  \setlength{\itemindent}{0pt}
  \setlength{\leftmargin}{0pt}
  \setlength{\parsep}{0pt}
  \ifodd #1
   \setlength{\leftmargin}{\cslhangindent}
   \setlength{\itemindent}{-1\cslhangindent}
  \fi
  \setlength{\itemsep}{#2\baselineskip}}}
 {\end{list}}
\usepackage{calc}

\ifLuaTeX
\usepackage[bidi=basic]{babel}
\else
\usepackage[bidi=default]{babel}
\fi
\babelprovide[main,import]{american}

\def\languageshorthands#1{}
\ifLuaTeX
  \usepackage{selnolig}  
\fi
\IfFileExists{bookmark.sty}{\usepackage{bookmark}}{\usepackage{hyperref}}
\IfFileExists{xurl.sty}{\usepackage{xurl}}{} 
\hypersetup{
  pdftitle={line racer: Rapid Calculation of Exoplanetary Radiative
Opacities},
  pdfauthor={David B. Hägele, Paul Mollière, Michiel Min},
  pdflang={en-US},
  colorlinks=true,
  linkcolor={Maroon},
  filecolor={Maroon},
  citecolor={Blue},
  urlcolor={Blue},
  pdfcreator={LaTeX via pandoc}}

\title{line racer: Rapid Calculation of Exoplanetary Radiative
Opacities}

\definecolor{c53baa1}{RGB}{83,186,161}
\definecolor{c202826}{RGB}{32,40,38}

\usepackage[affil-it]{authblk}
\usepackage{orcidlink}
\author[1,2%
  \ensuremath\mathparagraph]{David B. Hägele%
    \,\orcidlink{0009-0009-7667-7003}\,%
    }
\author[1%
  ]{Paul Mollière%
    \,\orcidlink{0000-0003-4096-7067}\,%
    }
\author[3%
  ]{Michiel Min%
    \,\orcidlink{0000-0001-5778-0376}\,%
    }

\affil[1]{Max-Planck-Institut für Astronomie, Königstuhl 17 D-69117
Heidelberg, Germany%
  }
\affil[2]{Ruprecht-Karls-Universität Heidelberg, Fakultät für Physik und
Astronomie, Im Neuenheimer Feld 226, 69120 Heidelberg, Germany%
  }
\affil[3]{SRON Space Research Organisation Netherlands, Niels Bohrweg 4,
2333 CA Leiden, The Netherlands%
  }
\affil[$\mathparagraph$]{Corresponding author: %
  haegele@mpia.de %
}
\date{}

\begin{document}
\maketitle

\section{Summary}\label{summary}

Detailed studies of exoplanet and brown dwarf atmospheres rely on
precise knowledge of the spectral features of possible atmospheric
species. These features are a result of the interaction of different
molecules and atoms with the radiation of the host star or the intrinsic
thermal radiation of the object. The interaction is described by the
opacity of a species, which determines how much light is absorbed at a
given wavelength. The strength and shape of these opacities are very
dependent on the temperature and pressure of the atmosphere, which is
why they have to be calculated for a wide range of conditions.
\texttt{line\ racer} is a Python package that enables computing
high-resolution opacities from large molecular and atomic line lists in
an effective manner. It offers users a wide range of options to
customize opacity calculations to their needs and available hardware.
The code is designed for efficient parallelization on multi-core and
multi-node systems and can produce outputs compatible with popular
atmospheric modeling and retrieval tools.

\section{Statement of need}\label{statement-of-need}

With the expected revolution in high-resolution spectroscopy of
exoplanet and brown dwarf atmospheres in the coming years, highly
accurate knowledge of the opacities of various molecular and atomic
species is crucial. The opacities are computed from line lists, which
contain the information about the transitions between different energy
levels of the molecules and atoms.

As a result, the line lists are growing rapidly in size. Databases like
ExoMol (\citeproc{ref-Tennyson2024}{Tennyson et al., 2024}) now have
line lists with billions of transitions for multiple molecules. For
example, the latest ExoMol methane line list MM
(\citeproc{ref-Yurchenko2024}{Yurchenko et al., 2024}) contains 50
billion lines. It is challenging to compute the opacities accurately
from these line lists in a reasonable time.

In general, \texttt{line\ racer} supports the line lists from ExoMol
(\citeproc{ref-Tennyson2024}{Tennyson et al., 2024}), HITRAN
(\citeproc{ref-HITRAN2024}{Gordon et al., 2026}), HITEMP
(\citeproc{ref-HITEMP2010}{Rothman et al., 2010}), and the new ExoAtom
(\citeproc{ref-ExoAtom}{Ni et al., 2026}) database, which uses NIST
(\citeproc{ref-NIST}{Ralchenko \& Kramida, 2020}) and Kurucz
(\citeproc{ref-Kurucz2011}{Kurucz, 2011},
\citeproc{ref-Kurucz2017}{2017}) data, enabling opacity calculations for
both molecular and atomic species. The code is explicitly designed to
calculate high-resolution opacities from large line lists in an
efficient manner. To achieve this, it uses a combination of two
different algorithms for the line profile calculation:

\begin{enumerate}
\def\labelenumi{\arabic{enumi}.}
\tightlist
\item
  A line profile calculation based on the algorithm by Humlíček
  (\citeproc{ref-Humlicek1982}{1982}) with a speed-up proposed by
  Mollière et al. (\citeproc{ref-Molliere2015}{2015}).
\item
  A sampling technique of the line profiles based on the algorithm
  proposed by Min (\citeproc{ref-Min2017}{2017}).
\end{enumerate}

This combination allows for a very fast calculation while maintaining a
high accuracy. Compared to existing tools, \texttt{line\ racer} has a
calculation time per transition that is dependent on the importance of
the transition, which is a significant advantage for the largest line
lists. Moreover, it is automatically parallelized to be used on multiple
nodes and cores, depending on the user's needs and available hardware.

Due to the increasing importance of observations at high spectral
resolution, it is also becoming relevant how exactly the line profiles
are treated in the opacity calculations. In particular, the line wing
treatment can have a significant impact on the resulting opacities at
high resolution. Therefore, \texttt{line\ racer} includes different
options for the line wing treatment, such as a cutoff at a
user-specified distance from the line center or a sub-Lorentzian
treatment following Hartmann et al. (\citeproc{ref-Hartmann2002}{2002})
or Burch et al. (\citeproc{ref-Burch1969}{1969}). More treatments will
be added over time or on user request.

The opacities calculated with \texttt{line\ racer} can be used for
various applications in the field of exoplanet and brown dwarf
atmospheres. To be directly compatible with one of the most used
atmospheric modeling and retrieval codes, \texttt{line\ racer} can
output the opacities in petitRADTRANS
(\citeproc{ref-Molliere2019}{Mollière et al., 2019}) format in addition
to a simple standard format.

\section{State of the field}\label{state-of-the-field}

There are several other tools to compute opacities for exoplanet
atmospheres, including
\href{https://helios-k2.readthedocs.io/en/latest/}{\texttt{HELIOS-K}}
(\citeproc{ref-Grimm2021}{Grimm et al., 2021}),
\href{https://cthulhu.readthedocs.io/en/latest/}{\texttt{Cthulhu}}
(\citeproc{ref-Agrawal2024}{Agrawal \& MacDonald, 2024}),
\href{https://exocross.readthedocs.io/en/latest/}{\texttt{ExoCross}}
(\citeproc{ref-Yurchenko2018}{Yurchenko et al., 2018}), and
\href{https://py-rox.readthedocs.io/en/latest/}{\texttt{pyROX}}
(\citeproc{ref-deRegt2025}{de Regt et al., 2025}). To our knowledge,
\texttt{line\ racer} is the first open source tool that has a
calculation time per transition that is dependent on the importance of
the transition. With this feature, line lists with more transitions do
not necessarily lead to longer calculation times, which is important
given how quickly these lists are growing in size. The underlying
advantages have been demonstrated by Min (\citeproc{ref-Min2017}{2017}).
We extended this approach, improved the robustness and performance, and
made it publicly available through \texttt{line\ racer}.

\section{Software design}\label{software-design}

The core line profile calculation algorithms are implemented in Fortran
to achieve high performance for these parts that are computationally
most expensive. The mentioned split between the two line profile
calculation algorithms was chosen to optimize the performance while
maintaining accuracy for all pressure and temperature combinations.
Through that, \texttt{line\ racer} is able to calculate opacities for
all line lists, even for future line lists of complex molecules that
will be even larger than the current ones. The overall package is
written in Python, which does the line list read-in, calculation of the
transition parameters, management of the parallelization, and the output
of the opacities. This enables an easy installation and accessibility
for a wide range of users. For the line profile calculation, the Python
code uses the Fortran algorithms. \texttt{line\ racer} includes
automatic tests using the
\href{https://tox.wiki/en/4.34.1/}{\texttt{tox}} framework to ensure
reliability and reproducibility of the code.

\section{Research impact statement}\label{research-impact-statement}

\texttt{line\ racer} is directly compatible with the widely used
atmospheric modeling and retrieval code petitRADTRANS
(\citeproc{ref-Molliere2019}{Mollière et al., 2019}). Opacities
calculated with \texttt{line\ racer} have already been used in
petitRADTRANS retrievals of directly imaged planets
(\citeproc{ref-Xuan2026a}{Xuan, Balmer, et al., 2026};
\citeproc{ref-Xuan2026b}{Xuan, Chachan, et al., 2026}). Additionally,
due to the flexible output in a standard format, it can be used for
other applications as well, and it is planned to include more output
formats in the future. Moreover, it will be used to update the opacities
database of petitRADTRANS to the latest available line lists and extend
the available high-resolution opacities. This enables improved forward
modeling and retrievals of exoplanet and brown dwarf atmospheres.

\section{Line profile calculation}\label{line-profile-calculation}

As introduced above, the line profile calculation in
\texttt{line\ racer} is automatically split up into two different
algorithms. The decision of which algorithm to use for the transitions
is based on their strength and the size of the line list in terms of the
total amount of transitions and their density.

For the strongest transitions of a large line list, or for small or
sparse line lists, the line profiles are calculated with the Humlíček
algorithm (\citeproc{ref-Humlicek1982}{Humlíček, 1982}) combined with a
speed-up proposed by Mollière et al.
(\citeproc{ref-Molliere2015}{2015}). The speed-up is achieved by
calculating the line wings on a coarser grid and interpolating them to
the fine grid.

The profiles of weaker transitions of large line lists are calculated
with a modified version of the line profile sampling technique proposed
by Min (\citeproc{ref-Min2017}{2017}). Instead of calculating the line
profiles on every wavelength grid point for every transition, it samples
them to a precision that is determined by the importance of the
transition. For example, if the transition is only contributing to the
continuum background of the opacity, just a few samples are used to
reproduce the overall opacity while minimizing the calculation time.

Atomic and molecular transitions have a shape described by the so-called
Voigt profile. A Voigt profile is a convolution of Lorentz and Doppler
profiles, which describe the effects of natural and pressure broadening
(Lorentz), and thermal broadening (Doppler). The Voigt sampling
technique is based on the idea of sampling the Lorentz and Doppler
profiles separately for each Voigt profile sample. To get the location
of the Voigt sample that contributes to the overall opacity, the Lorentz
and Doppler positions are added to the line center position. In Min
(\citeproc{ref-Min2017}{2017}), the weight of the sampled contribution
that is added to the total opacity is determined by the sampled Lorentz
shift \(\Delta\nu_\mathrm{press}\), among other things. More concretely,
Min (\citeproc{ref-Min2017}{2017}) defined sample-dependent weights
\(w\) that reproduce the shape of the line profile. However, the
wavelength integral of the line profile needs to be calculated to
normalize the line to the correct transition intensity \(S\). The
normalization afterward is time-consuming, which is why we introduced a
new form of this weight that ensures that the integrated intensity is
directly correct.

\[
w = \frac{2\Delta\nu_\mathrm{press}^2}{\Delta\nu_\mathrm{press}^2 + \gamma^2}\frac{S}{\mathrm{N_{samples}}}\frac{R}{\nu_\mathrm{eff}}
\]

The first term is adapted from Min (\citeproc{ref-Min2017}{2017}) to
preserve the shape of the line profile and includes the pressure
broadening width of the line profile \(\gamma\). The other terms
represent the contribution of the intensity \(S\), the number of samples
\(\mathrm{N_{samples}}\), the resolution \(R\), and the effective
wavenumber \(\nu_\mathrm{eff}\) of the transition.

The direct normalization is introducing a small uncertainty in the
integrated intensity because the weight and therefore the contribution
are dependent on the random samples. But since it is only used for dense
line lists and scales inversely with the number of transitions and
samples, it is negligible.

In addition to the improved weight, we also enhanced the number of
samples for each transition, increasing precision overall, and
especially for low-pressure cases. More details can be found in the
documentation of the code, together with benchmark results and timing
information.

\section{AI usage disclosure}\label{ai-usage-disclosure}

AI was used only for minor assistance in developing
\texttt{line\ racer}, including support with debugging code, assistance
with documentation, and proofreading an earlier draft of the paper for
grammatical and stylistic improvements. Moreover, every output generated
by AI was carefully reviewed and edited. The core algorithms, software
design and scientific content were developed independently by the
authors.

\section{Acknowledgements}\label{acknowledgements}

We thank Jerry Xuan for testing the early versions of this code and
providing valuable feedback to the code and the documentation.

\section*{References}\label{references}
\addcontentsline{toc}{section}{References}

\protect\phantomsection\label{refs}

\end{document}